# What Are You Hiding? Algorithmic Transparency and User Perceptions

Aaron Springer, Steve Whittaker

University of California, Santa Cruz, 1156 High St., Santa Cruz, California, USA
{alspring, swhittak}@ucsc.edu

**Abstract**

Extensive recent media focus has been directed towards the dark side of intelligent systems, how algorithms can influence society negatively. Often, transparency is proposed as a solution or step in the right direction. Unfortunately, research is mixed on the impact of transparency on the user experience. We examine transparency in the context an interactive system that predicts positive/negative emotion from a users' written text. We unify seemingly this contradictory research under a single model. We show that transparency can negatively affect accuracy perceptions for users whose expectations were not violated by the system's prediction; however, transparency also limits the damage done when users' expectations are violated by system predictions.

## Introduction

Intelligent systems powered by machine learning are pervasive in our everyday lives. These systems make decisions ranging from the mundane to the magnificent, from routes to work to recommendations about criminal recidivism. We, as humans, increasingly devolve more and more responsibility to these systems with little transparency or oversight. Concerns about how these systems are making decisions are building and this is only exacerbated by recent machine learning methods such as deep learning that are difficult to explain in human-comprehensible turns. These opaque systems have taken blame for major events such as the 2016 election of Donald Trump (Olson, 2016); they have been implicated in disadvantaging minority prisoners that are up for parole (Julia Angwin, 2016). Life-changing decisions are being made without any ability to examine the method or data these algorithms are using to make predictions.

This lack of transparency enables algorithmic problems to run amok. These systems have been shown to capture societal and human biases and perpetuate them systemically (Bolukbasi, Chang, Zou, Saligrama, & Kalai, 2016). Other work has indicated that a lack of transparency may lead users to accept output from algorithms that are simply random (Springer, Hollis, & Whittaker, 2017). It seems that we are seeing more automation bias than ever, (Cummings, 2004) we are increasingly willing to go along

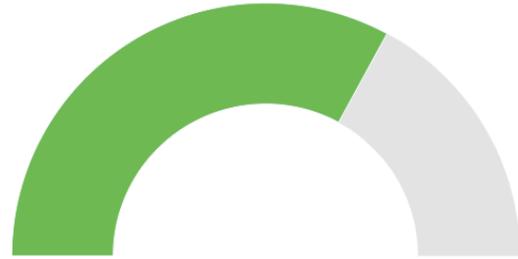

*Figure 1. The E-meter System in the Transparent Condition after a user wrote about a positive experience and the E-meter predicted mood and associated words accurately*

with what systems suggest rather than trying to critically examine those suggestions.

All of these problems have been met by calls for industry implementation of transparent algorithms and expanded research into issues of transparency and trust in algorithms. The response from industry has been anemic. Few commercial products have accepted this challenge of increased transparency. Yes, algorithms are evolving quickly, but the bigger issue seems to be that methods for transparency are not well understood. Results around the effects of algorithmic transparency have been mixed. Lim and Dey (Lim & Dey, 2011) found that increased transparency can make users question the algorithm when it's correct, therefore impairing the user experience. Users may also feel that these explanations simply cause additional processing without offering real value (Bunt, Lount, & Lauzon, 2012). On the other hand, transparency can help protect system trust by allowing users to understand why a prediction was made when that prediction violates their expectations (Kizilcec, 2016). It is difficult to form a coherent picture of the effects of transparency on the user experience from these conflicting results.

In this study, we explore the effects of transparency on the user experience. We experiment in the context of the E-

meter, a system that predicts positive/negative emotion in a user's account of a past experience. The E-meter context allows us to examine how users interact with a system making predictions in an area the user is an expert in; only the user has the ground truth of their emotions. We make the E-meter more transparent to users to examine the conflicting nature of previous transparency research.

Specifically, we focus on how adding transparency to the E-meter influences user perceptions of accuracy. Accuracy is an important aspect of user experiences with intelligent systems. Users who believe that a system is accurate may be more likely to act upon its recommendations (Hollis et al., 2017). Therefore, transparency could play an important role in motivating user engagement with intelligent health and mental health applications. However, since recent research on transparency has mixed results, implementing transparency could also have net negative affects and push users away from using the application. We must unify these conflicting results in a way that illuminates a path forward for the use of transparency in intelligent systems.

## Methods

### Users

Users were recruited from Amazon Turk and paid $3.33 to evaluate the E-meter system. This evaluation took 13 minutes on average. We recruited 41 users to test the E-meter system across two conditions who were screened for stable mental health. Users were divided into 2 conditions: a control condition, and a transparent condition that allowed users to examine how each word affected the E-meter overall.

### Machine Learning Model

Emotional valence predictions for users' experiences were predicted using a linear regression model trained on text from the EmotiCal project (Hollis et al., 2017). In EmotiCal, users wrote short textual entries and logged their overall mood, which gave us a supervised training set to train our linear regression on. We trained the linear regression on 6249 textual entries and mood scores from 164 EmotiCal users. Text features were stemmed using the Porter stemming algorithm (Porter, 1980) and then the top 600 unigrams were selected by f-score. Using a train/test split of 85/15 the linear regression tested at $R^2 = 0.25$; mean absolute error was .95 on the target variable (mood) scale of (-3,3). In order to implement this model on a larger range for the E-meter, we scaled the predictions to (0,100) to create a more continuous and variable experience for users. The mean absolute error of our model indicates that the E-meter will, on average, err by 15.83 points on a (0,100) scale for each user's mood prediction.

### E-meter System

The E-meter (Figure 1) presented users with a web page showing a figure, a short description of the system, instructions, and a text box to write in. The system was described as an "algorithm that assesses the positivity/negativity of [their] writing". The instructions asked users to "Please write at least 100 words about an emotional experience that affected you in the last week."

As users wrote, the E-meter moved in accordance with the emotional valence of their writing; the meter could move positively, towards filling the gauge to the right, or negatively, towards emptying the gauge to the left, based on a regression model predicting the mood of their written experience. The E-meter was updated in real time after the user finished writing or removing a new word in the text box. The color of the E-meter changed depending on how positive or negative the overall rating, the E-meter changed from a deep red for very negative ratings, through orange, yellow, and light green, all the way to a dark green for very positive ratings of the user's' text.

The E-meter randomly assigned users to either a transparent or control condition. Those in the transparent condition were told that individual words would be highlighted to show the word's contribution to the E-meter's overall rating of their affect. In the transparent condition, users were able to see the extent to which each individual word they wrote contributed to the overall E-meter rating using a method that highlighted words according to their evaluated affect. Users in the control condition did not see their words highlighted, though they could still see the movement of the meter after they finished writing each word.

We operationalized transparency in this space by passively highlighting the valence association of each word in the model. This form of transparency offers a view directly through the text to the regression model powering the E-meter; it portrays how strongly the regression model correlates each word with positive or negative emotion. Offering this persistence of transparency allows users to reexamine what they had written when they finished and reconcile their overall E-meter rating with the fine-grained transparency from the text. When expectations are violated, users are prone to seek out more information to learn why the violation happened (Kizilcec, 2016). Our operationalization of transparency allows users to engage in this questioning mode.

### Survey

Following their experimentation with the E-meter, users were asked various questions about their experiences. Importantly, we asked users their perceptions of their own writing and their perception of the E-meter's rating of their writing on a 7-point Likert scale from "Strongly Negative" to "Strongly Positive". Users were additionally asked about the accuracy of the E-meter rating (7 point, "Very Inaccurate" to "Very Accurate") and how trustworthy they found the system (5 point, "Not at all" to "Extremely" trustworthy) and reasons for these ratings. The final questions were open-ended and asked about users' likes/dislikes and their' theories about how the system was calculating their final score.

|  | Coefficient | Std. Error | p-value |
|---|---|---|---|
| Intercept | 6.991 | 0.377 | **< 0.00001** |
| Expectation Violation | -1.736 | 0.601 | **< 0.00001** |
| Transparency | -1.406 | 0.198 | **0.007** |
| Transparency * Expectation Violation | 1.057 | 0.441 | **0.022** |

*Table 1. Linear Regression Predicting Users' Accuracy Perceptions*

## Results

The majority of users across conditions found the E-meter to be "Accurate" or "Very Accurate" with the median being "Accurate". Users were slightly less trusting of the meter and found it to be "Moderately Trustworthy".

### Transparency Moderates Expectation Violation

We calculate a user's expectation violation of their overall rating by subtracting the user's perception of their own writing (their expectation of the E-meter value if it were perfect) from the actual perception of the final E-meter score. If a user felt that their writing was "Strongly Negative" (1) but the E-meter rated it as "Slightly Negative" (3) then the user's expectation violation would be 2. Therefore, higher levels of expectation violation indicate that the user felt that the E-meter was less accurate overall while low levels of expectation violation should correlate with increased perceptions of accuracy. We refer to accuracy from the survey as holistic accuracy of the system, encompassing perceptions of the meter and the word highlighting. We see a strong relationship between expectation violation and accuracy in the control group, $r = -.898$, $p < .00001$ as well as in data from our previous study (Springer et al., 2017). Interestingly, this correlation between expectation violation and accuracy perceptions disappears in the transparent condition: $r = -0.175$, $p = 0.488$. We find that the relationship between expectation violation and accuracy perceptions is not so simple in the presence of transparency.

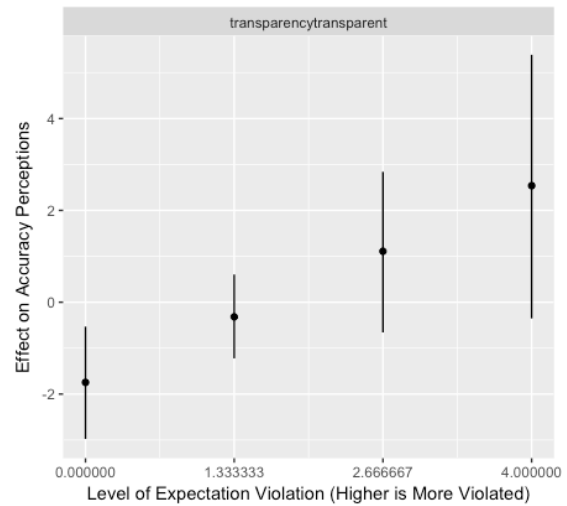

*Figure 2. Transparency and Expectation Violation Interaction*

We find that transparency and expectation violation interact complexly. We modeled this interaction using a linear regression predicting user accuracy perceptions (see Table 1). The regression was highly predictive $R^2=.548$, $p < .000001$. Transparency actually has a net negative effect on perceptions of accuracy. However, transparency begins to have a positive effect in the presence of increased expectation violation.

In the control condition expectation violation resulted in decreased perceptions of system accuracy. Users of the transparent system saw less decrease in accuracy perceptions as expectation violation increased. However, transparent system users have lower perceptions of the E-meter's accuracy even when their expectations are not violated.

### Qualitative Examination of Anomalous Users

We now turn to qualitatively examining users to discern how transparency could cause this decrease in accuracy perceptions. We specifically examine these instances where the user stated that the E-meter was within 1 point of the user's own perception of their writing (on the 7-point scale from 'Strongly Negative' to 'Strongly Positive') and the user still rated the accuracy as inaccurate.

**Lack of Personalization for Users**

One problem concerned the generalized nature of the mood models. The models were trained on data consisting of 164 different users and thus the model learned general associations that hold across most people. Of course specific individuals may have completely different associations for specific words and making the machine learning transparent by highlighting these words can expose these differences more readily (Hollis et al., 2017). One user wrote: "Family is bad for me but it was marked in green."

This user felt that his specific family experiences were highly negative in contrast with the model's association.

**Questions Created by Transparency**

The goal of highlighting words and making the E-meter calculation more transparent was to passively explain to users where their final positivity/negativity rating was coming from. However, for some users, this transparency just created more questions. One user noted that their final negative rating didn't make sense "because the rating did not correspond to the number of identified words". Other users wanted to know how the ratings the highlighted words were originally identified and their association with mood calculated. Certain presentations of transparency may be more likely to evoke these doubting questions.

## Conclusion

Algorithmic transparency has a complex relationship with user perceptions of algorithmic accuracy. In our experiment, transparency effectively compressed user perceptions of accuracy. Transparency users with the most violated expectations had better perceptions of the E-meter's accuracy compared to their control counterparts. However, users in the transparent condition were less likely to regard the E-meter as highly accurate even when it did not violate their expectations at all. This result unifies seemingly contradictory results that indicated that transparency had positive or negative effects (Kizilcec, 2016; Lim & Dey, 2011).

Whether or not transparency is a net positive in an application may depend on other characteristics of the application. For example, if an intelligent system is highly accurate in its predictions, then increasing the transparency of the application may have a net negative effect of lowering perceptions of accuracy. If an application is inaccurate, then transparency could have a net positive by tempering those negative accuracy perceptions. Of course, other factors influence whether transparency is needed, such as the impact of the decision that the algorithm is influencing.

In addition, we find a few routes through which transparency can decrease perceptions of accuracy. Transparency exposes the fact that these models are general and learned from a societal space that embodies many correlations that may not be personally relevant to each user. These correlations can be learned over time, like in EmotiCal (Hollis et al., 2017), but on first interaction this is a difficult problem to solve. Possible solutions could include "emotional onboarding" analogous to what recommender systems use to solve the cold start problem, others may include scanning a user's social media and building a profile for them through the given information (Warshaw et al., 2015). Another problem stemmed from the exposed information from transparency creating additional questions which led to users doubting system accuracy. Overall, we want a seamful design allowing users to explore the model but only to the point that is helpful.

Rather than simply focusing on how to present transparency in ways that don't evoke more questions of the algorithm, researchers should instead focus on recognizing expectation violation. If a system can recognize in real time when a users' expectations of the system were violated, it can choose to be selectively transparent. Selective transparency would maintain the positive aspects of transparency without making users unduly question the application.